\documentclass[12pt,preprint]{emulateapj}

\usepackage{natbib}
\def\degree{\ifmmode {^\circ}\else {$^\circ$}\fi}
\def\rstar{\ifmmode {\, R_{\star}}\else $R_{\star}$\fi}
\def\msol{\ifmmode {\, M_{\odot}}\else $M_{\odot}$\fi}
\def\rsol{\ifmmode {\, R_{\odot}}\else $R_{\odot}$\fi}
\def\lsol{\ifmmode {\, L_{\odot}}\else $L_{\odot}$\fi}
\def\msolyr{\ifmmode {\, M_{\odot}\,{\rm yr}^{-1}}\else $M_{\odot}\,{\rm yr}^{-1}$\fi}
\def\mdot{\ifmmode {\,\dot{M}}\else $\dot{M}$\fi}
\def\mdotyr{\ifmmode {\,\dot{M}\,yr^{-1}}\else $\dot{M}\,yr^{-1}$\fi}

\newcommand{\Teff}{\ifmmode{T_{\rm eff}}\else{$T_{\rm eff}$}}

\begin{document}

\title{Herschel Observations of a Newly Discovered UX Ori Star in the Large Magellanic Cloud}

\author{Geoffrey C. Clayton\altaffilmark{1}, B. Sargent\altaffilmark{2}, M.L. Boyer\altaffilmark{2}, B. A. Whitney\altaffilmark{3}, Jacco Th. van Loon\altaffilmark{4}, M. Meixner\altaffilmark{2}, P. Tisserand\altaffilmark{5} 
C. Engelbracht\altaffilmark{6}, S. Hony\altaffilmark{7}, R. Indebetouw\altaffilmark{8}, K. A. Misselt\altaffilmark{5}, K. Okumura\altaffilmark{6}, P. Panuzzo\altaffilmark{6}, J. Roman-Duval\altaffilmark{2}, M. Sauvage\altaffilmark{6}, J.M. Oliveira\altaffilmark{6}, M. Sewilo\altaffilmark{2}, and E. Churchwell\altaffilmark{9}}

\altaffiltext{1}{Department of Physics \& Astronomy, Louisiana State
University, Baton Rouge, LA 70803; gclayton@fenway.phys.lsu.edu}

\altaffiltext{2}{Space Telescope Science Institute, 3700 San Martin Drive,
Baltimore, MD 21218, USA; duval, gordon, mboyer, meixner, sargent@stsci.edu}

\altaffiltext{3}{Space Science Institute, 4750 Walnut St. Suite 205, Boulder, CO
80301, USA; bwhitney@spacescience.org}

\altaffiltext{4}{School of Physical \& Geographical Sciences, Lennard-Jones
Laboratories, Keele University, Staffordshire ST5 5BG, UK; jacco, joana@astro.keele.ac.uk}

\altaffiltext{5}{ANU-RSAA, Mount Stromlo Observatory, Cotter Road, Weston Creek
ACT 2611, Australia; tisseran@mso.anu.edu.au}

\altaffiltext{6}{Steward Observatory, University of Arizona, 933 North Cherry
Ave., Tucson, AZ 85721, USA; cengelbracht, misselt@as.arizona.edu}

\altaffiltext{7}{CEA, Laboratoire AIM, Irfu/SAp, Orme des Merisiers, F-91191
Gif-sur-Yvette, France; koryo.okumura, marc.sauvage, pasquale.panuzzo, sacha.hony@cea.fr}

\altaffiltext{8}{National Radio Astronomy Observatory, Department of
Astronomy, University of Virginia, PO Box 3818, Charlottesville,
VA 22903 USA; remy@virginia.edu}

\altaffiltext{9}{Department of Astronomy, 475 North Charter St., University of
Wisconsin, Madison, WI 53706, USA; ebc@astro.wisc.edu}



\begin{abstract}
The LMC star, SSTISAGE1C J050756.44--703453.9, was first noticed during a survey of EROS-2 lightcurves for stars with large irregular brightness variations typical of the R Coronae Borealis (RCB) class. However, the visible spectrum showing emission lines including the Balmer and Paschen series as well as many Fe II lines is emphatically not that of an RCB star. 
This star has all of the characteristics of a typical UX Ori star. It has a spectral type of approximately A2 and has excited an H II region in its vicinity. However, if it is an LMC member, then it is very luminous for a Herbig Ae/Be star. It shows irregular drops in brightness of up to 2 mag, and displays the reddening and ``blueing" typical of this class of stars. Its spectrum, showing a combination of emission and absorption lines, is typical of a UX Ori star that is in a decline caused by obscuration from the circumstellar dust. SSTISAGE1C J050756.44--703453.9 has a strong IR excess and significant emission is present out to 500 \micron. Monte Carlo radiative transfer modeling of the SED requires that SSTISAGE1C J050756.44--703453.9 has both a dusty disk as well as a large extended diffuse envelope to fit both the mid- and far-IR dust emission.
This star is a new member of the UX Ori subclass of the Herbig Ae/Be stars and only the second such star to be discovered in the LMC.

\end{abstract}



\section{Introduction}
The Herbig Ae/Be stars are A or B stars with emission line spectra located in dusty star forming regions and which are typically exciting a small emission nebula nearby  \citep{1960ApJS....4..337H}. In addition, these stars generally show IR excesses due to circumstellar dust and have luminosity classes III - V \citep{1998ARA&A..36..233W}. 
The UX Ori subclass of the Herbig Ae/Be stars is characterized by sudden declines of up to 3 mag at V, associated with increased extinction and polarization \citep{1998ARA&A..36..233W}.  These stars often show spectroscopic variability as well. All of these variations have been ascribed to changes in a dusty disk viewed nearly edge-on. 
Suggested scenarios for the declines in brightness are infalling proto-cometary objects, eclipses by optically thick clouds in an optically thin disk or self-shadowed flared disks \citep{2002ApJ...564..405R,2003ApJ...594L..47D,2000prpl.conf..613G,2000A&A...363..984B}.
\citet{2000A&A...364..633N} suggest that the disk doesn't have to be exactly edge-on but that declines can occur for a range of inclinations. 

\citet{1994AJ....108.1906H} identify 18 Galactic stars that may belong to the UX Ori class.  
About half of the Herbig Ae/Be stars later than B9 are UX Ori stars
\citep{1997ApJ...491..885N}.
A number of candidate pre-main-sequence (PMS) stars have been identified in the LMC \citep[e.g.,][]{1996Sci...272..995B,2001AJ....122..858B,2009ApJ...699..150S}.
\citet{2002A&A...395..829D} found 21 LMC candidates after searching a sample of 80,000 variable stars in the EROS sample. 
These stars were picked as candidates because they were blue, had irregular brightness variations and hydrogen emission lines.
The LMC stars seem to be more luminous than their Galactic Herbig Ae/Be star counterparts \citep{2001AJ....122..858B,2002A&A...395..829D}. 
One star in this  sample, ELHC7, displays the attributes of a UX Ori star \citep{1999A&A...341..827L}. It is the only star in the sample with  a significant NIR excess and irregular sharp brightness drops \citep{2005A&A...432..619D}. ELHC7 also appears in the YSO candidate list developed by SAGE (Spitzer Survey of the Large Magellanic Cloud: Surveying the Agents of a Galaxy's Evolution)  \citep{2008AJ....136...18W}.
One of seven Herbig Ae/Be star candidates in the SMC, ESHC7 has also shown UX Ori characteristics  \citep{2003A&A...410..199D}.

The EROS-2 star, lm056-0K-21075 (also MACHO-9.4391.25), was found to be variable with irregular declines reminiscent of an R Coronae Borealis (RCB) star \citep{2009A&A...501..985T}. However, its visible spectrum is rich in hydrogen lines so this star is not a hydrogen-deficient RCB  star.
It is possibly the same emission-line star (No.\ 129) discovered in an H$\alpha$ survey by \citet{1963IrAJ....6..127L}. 
This star also appears as No.\ 288 (SSTISAGE1C J050756.44--703453.9) in the SAGE list of $\sim$1000 YSO candidates \citep{2008AJ....136...18W}. 
It was also identified as a YSO using data obtained by the HERschel Inventory
of The Agents of Galaxy Evolution (HERITAGE) Legacy program \citep{2010A&A...518L..73S}.
Hereafter, we will refer to this star as SAGE050756--703453. In this paper, we will argue that it is a UX Ori star which is also a member of the LMC. 

\begin{figure*}
\figurenum{1} 
\begin{center}
\includegraphics[width=6in,angle=0]{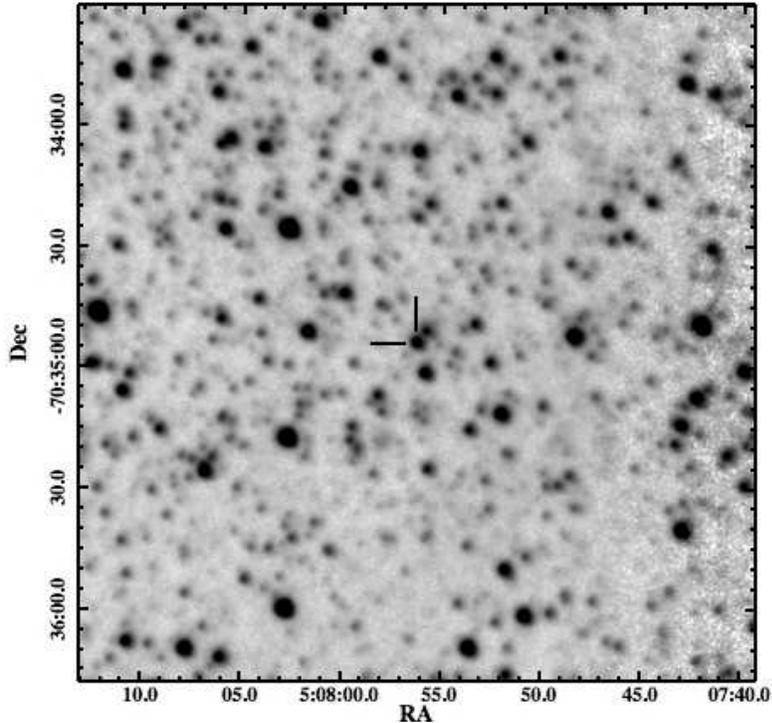}
\end{center}
\caption{Red EROS-2 image with the location of SAGE050756--703453 marked. North is up and east is to the left.}
\end{figure*}

\section{Observations and Data Reduction}

SAGE050756--703453 (2MASS J05075645-7034537) lies at RA = 05$^h$ 07$^m$ 56\fs45, and Dec =
$-$70\arcdeg~34\arcmin~53\farcs7 \citep[J2000;][]{2003tmc..book.....C}. A finding chart is shown in Figure 1.

Figure 2 shows the visible spectrum of SAGE050756--703453, obtained on JD 2454510.0 with the 
Dual-Beam Spectrograph (DBS) \citep{1988PASP..100..626R} attached
to the ANU 2.3m telescope at Siding Spring
Observatory \citep{2009A&A...501..985T}. The visible
wavelength band is split by a dichroic at around 6000 \AA\ and feeds
two spectrographs, with red and blue optimized
detectors. The full slit length is 6\farcm7. The spectrum has a 2-pixel resolution of 2 \AA.
The spectrum has been normalized. 

Figure 3 shows the lightcurve for SAGE050756--703453 including data from both the MACHO and EROS-2 projects.  The MACHO V and R magnitudes have been transformed to Johnson-Cousins V and R 
\citep{1999PASP..111.1539A}. The EROS-2 blue and red filters are very broad and not easily transformed. The B$_E$ and R$_E$ magnitudes are calibrated as described in \citet{2009A&A...501..985T}. The bands are centered between V and R, and between R and I, respectively. The B$_E$ - R$_E$ can be transformed to Cousins V-I (=1.02(B$_E$ - R$_E$)  \citep{1997A&A...318L..47B}.

Figure 4 shows a montage of IR images of the SAGE050756--703453 field including Spitzer/MIPS 24 \micron, and the Herschel PACS 100 and 160 \micron, and SPIRE, 250, 350, and 500 \micron~images.
The available photometry is summarized in Table 1. 
 The star is clearly visible as a point source in the center of each field.
The UBVRI photometry is from \citet{2002ApJS..141...81M} and \citet{2004AJ....128.1606Z}. The JHK photometry is from 2MASS  \citep{2003tmc..book.....C}. 
Herschel data were obtained as part of the HERITAGE Science Demonstration Program. PACS (100 and 160 \micron) and SPIRE (250, 350, and 500 \micron)
fluxes were extracted using aperture photometry \citep{2010A&A...518L..71M}. Apertures were
roughly the size of the full-width half-maximum of the PSF
and large sky apertures were chosen to avoid regions
of high IR background. The aperture corrections for PACS and
SPIRE were estimated using the currently available point spread
functions (PSFs).  To complement these data,
we have re-examined the SAGE MIPS images (24, 70, and 160 \micron), with
aperture corrections from the IRAC and MIPS data handbooks \citep{2006AJ....132.2268M}. 
The MIPS photometry was redone to ensure that the stellar and sky apertures matched those used for the Herschel images but the resulting photometry is consistent with the SAGE photometry \citep{2008AJ....136...18W}.
The SAGE photometry was used for the IRAC bands. We note that SAGE050756--703453 was not identified as an infrared variable in the comparison of the two SAGE survey epochs as there was no significant difference between the two measured fluxes  \citep{2009AJ....137.3139V}.
SAGE050756--703453 was not detected by IRAS. 

\begin{deluxetable}{cl}
\tablecaption{SAGE050756--703453 Photometry}
\tablenum{1}
\tablehead{\colhead{Band$^a$}&
           \colhead{Flux (Jy)}}
\startdata

U &  4.18e-04  $\pm$ 2.60e-05\\
& 6.56e-05  $\pm$ 3.81e-06\\
 B &  9.91e-04   $\pm$   4.90e-05\\
 & 1.15e-04 $\pm$  5.61e-06\\
 V  & 1.51e-03   $\pm$   7.00e-05\\
 & 1.99e-04 $\pm$  9.00e-06\\
 R & 1.82e-03   $\pm$   1.45e-04\\
 I & 1.05e-03 $\pm$  8.87e-05\\
 2MASS/J  & 1.02e-03   $\pm$   6.13e-05\\
 2MASS/H &  2.88e-03    $\pm$  1.54e-04\\
 2MASS/K  & 6.92e-03   $\pm$   1.98e-04\\
IRAC/3.6 &  2.06e-02   $\pm$   5.14e-04\\
IRAC/4.5 &  2.53e-02   $\pm$   1.10e-03\\
IRAC/5.8  & 3.07e-02   $\pm$   9.35e-04\\
IRAC/8.0 &  3.41e-02   $\pm$   7.54e-04\\
MIPS/24 &  6.36e-02   $\pm$   2.45e-04\\
MIPS/70 &  4.05e-01    $\pm$  1.32e-02\\
MIPS/160 &  9.07e-01   $\pm$   1.71e-01\\
PACS/100 &  4.73e-01   $\pm$   1.40e-02\\
PACS/160 &  6.33e-01   $\pm$   2.80e-02\\
SPIRE/250  & 4.88e-01   $\pm$   2.22e-02\\
SPIRE/350 &  4.03e-01   $\pm$   1.69e-02\\
SPIRE/500 &  1.37e-01   $\pm$   1.65e-02\\
\enddata
\tablenotetext{a}{There are two epochs of UBVRI photometry from \citet{2002ApJS..141...81M} and \citet{2004AJ....128.1606Z}. }
\end{deluxetable}

Figure 5 shows the spectral energy distribution (SED) of SAGE050756--703453. While Figure 5 gives a general idea of the SED, it should be remembered that it is a variable star and that the data from the various instruments included were taken at epochs that are often years apart. 
For instance, \citet{2002ApJS..141...81M} reports V= 16.00$\pm$0.05, B-V=0.55$\pm$0.05, U-B=0.05$\pm$0.07, V-R= 0.43$\pm$0.09, and \citet{2004AJ....128.1606Z} reports V= 18.21$\pm$0.05, B-V=0.72$\pm$0.05, U-B=-0.27$\pm$0.07, V-I= 2.25$\pm$0.09 for observations separated by several years. 



\section{Discussion}

The visible spectrum of SAGE050756--703453, plotted in Figure 2, shows P-Cygni profiles in the lower Balmer lines. The rest of the Balmer series is clearly visible in absorption. Other lines are mostly in emission including many Fe II lines in the 4500-5500 \AA\ region, the Ca II IR triplet, and the Paschen series. The blueshifted absorption in H$\alpha$ extends to $\sim$450 km s$^{-1}$. There was no photometric coverage of the star when the spectrum was obtained, but it is very similar to a spectrum of the UX Ori star, RR Tau (spectral type A2), obtained on 1998 Feb 14, when the star was in a decline $\Delta$V= 1.7 mag below maximum brightness
\citep{2002ApJ...564..405R}. The emission lines become visible in declines when the dust cloud eclipses the photosphere of the star but not the surrounding emission region. 
The line ratios of the Ca II IR triplet are close to 1:1:1 in both stars indicating that the gas is optically thick. 
The wings of the Balmer lines in RR Tau seem to be broader than those in SAGE050756--703453 possibly indicating that RR Tau may be less luminous \citep{2002ApJ...564..405R}. Another indication that SAGE050756--703453 is luminous is the strength of the O I $\lambda$$\lambda$7774, 8446 triplets \citep{1971BOTT....6..137M}. O I $\lambda$$\lambda$7774 is in absorption in SAGE050756--703453, while O I $\lambda$8446 is in emission in both stars.  The strength of the O I $\lambda$8446 is due to a fluorescence with Ly$\beta$ \citep{1947PASP...59..196B}.

Because SAGE050756--703453 was in a decline when the spectrum was taken and many lines are in emission, an accurate spectral type cannot be determined \citep{2004AJ....127.1682H}. However, the lines in the blue are in absorption and show no sign of emission cores. 
The line of He I $\lambda$4471 is weak or absent, but there is a significant Ca II K absorption line present \citep{1985A&A...151..340F}. Also, the Balmer lines are strongly in absorption and there is no sign of the G band. So based on these few lines, SAGE050756--703453 is an A star \citep{2004AJ....127.1682H}. The similarity of the spectrum to that of RR Tau and the absorption line strengths indicate that the best guess spectral type is A2\footnote{A Digital Spectral Classification Atlas by R. O. Gray, http://nedwww.ipac.caltech.edu/level5/Gray/frames.html}.  There are no atomic or molecular features of carbon so this is not a dust producing carbon star. 

\begin{figure*}
\figurenum{2} 
\begin{center}
\includegraphics[width=5in,angle=0]{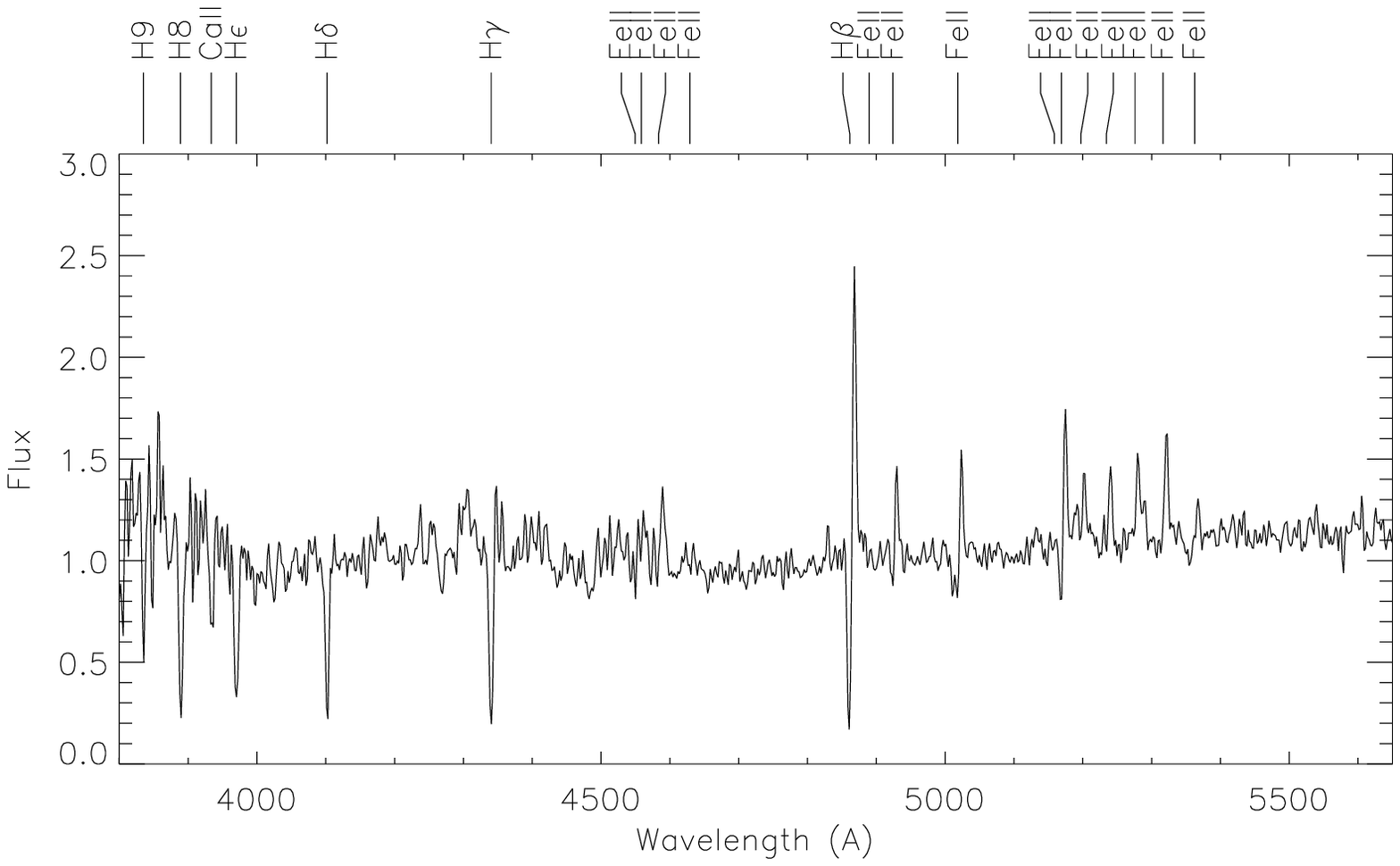}\\
\includegraphics[width=5in,angle=0]{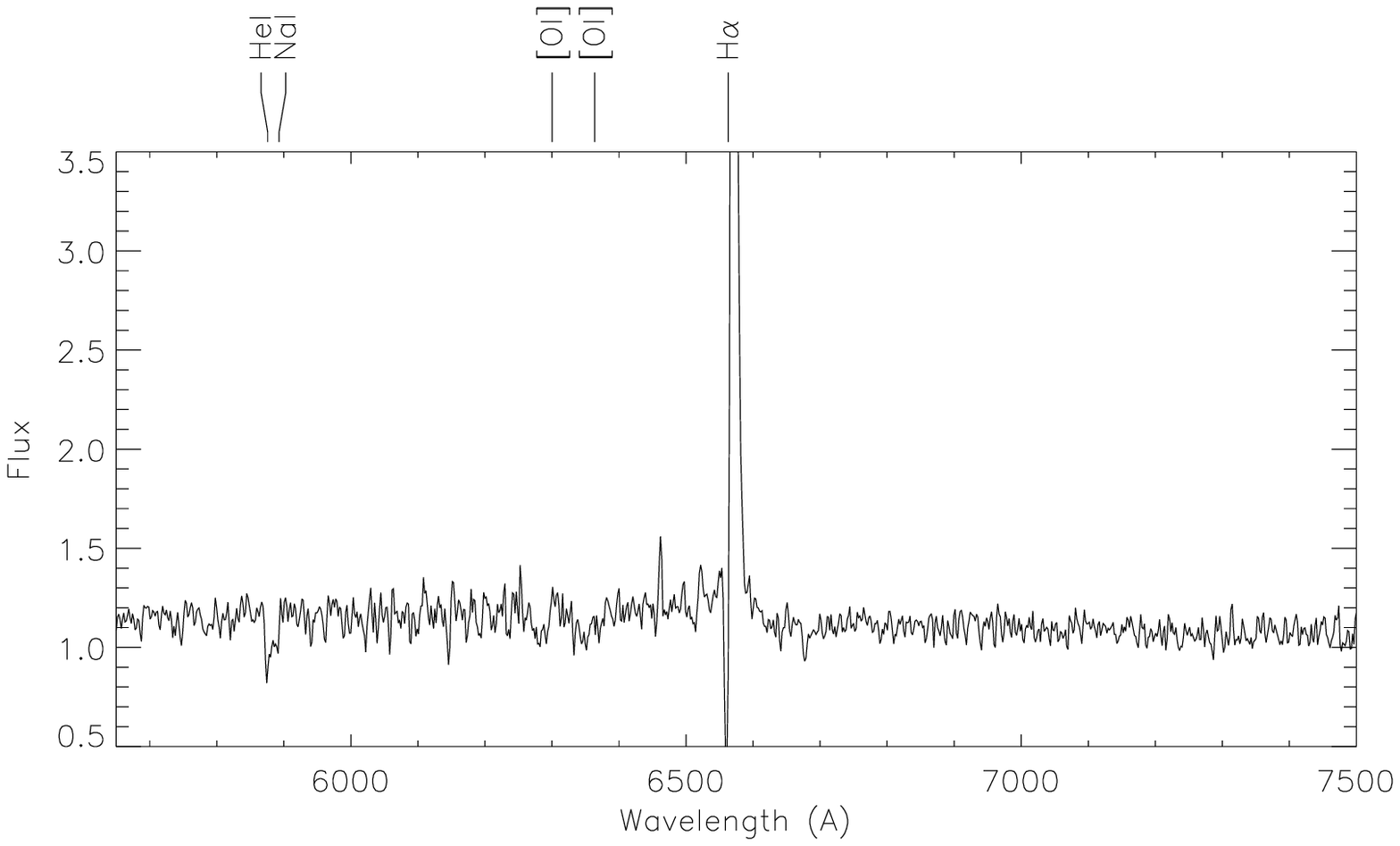}\\
\includegraphics[width=5in,angle=0]{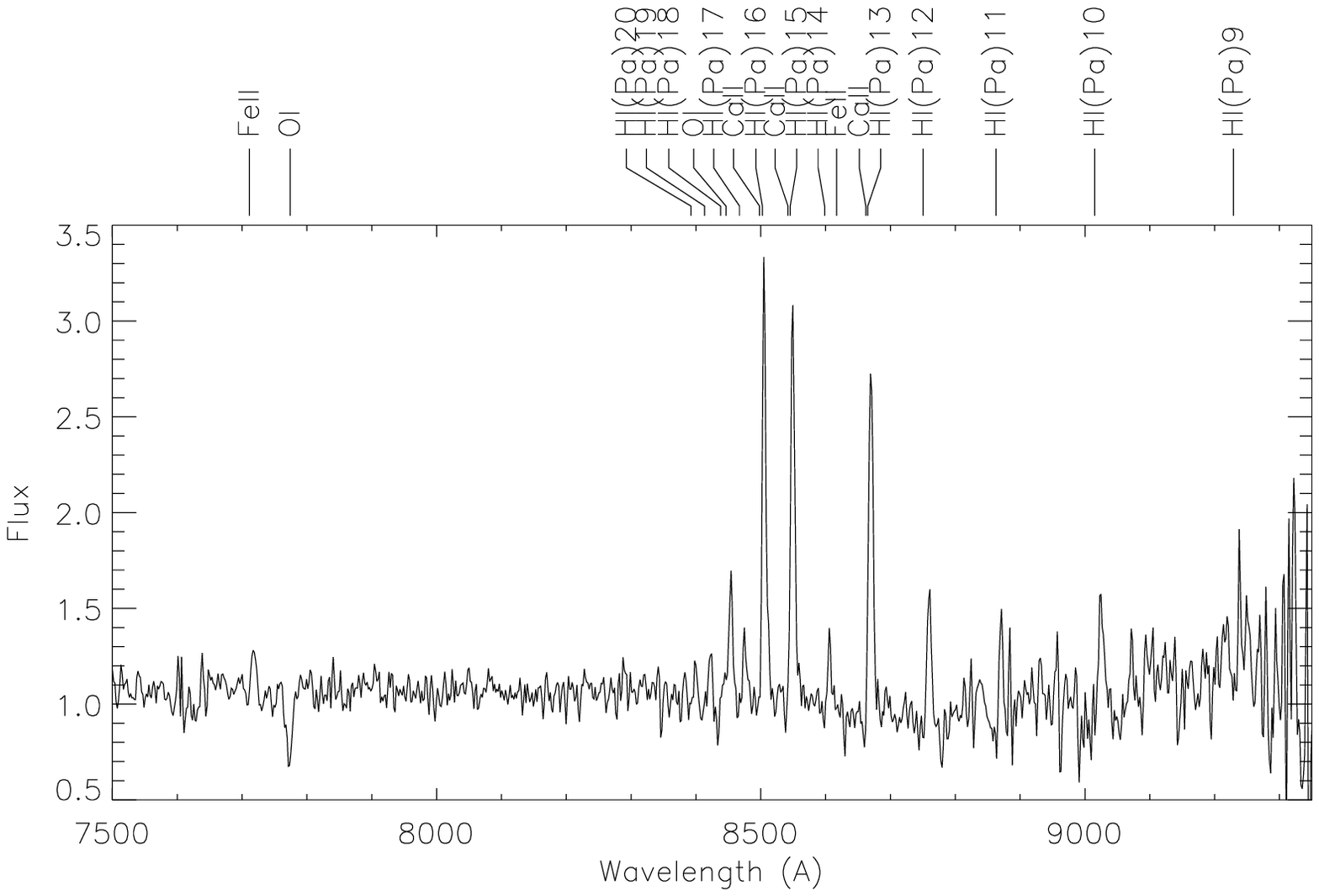}
\end{center}
\caption{The visible spectrum of SAGE050756--703453 with the prominent emission and absorption  lines identified. The spectrum has been normalized.}
\end{figure*}

The MACHO and EROS-2 lightcurves for SAGE050756--703453, shown in Figure 3, cover almost 11 yr. Unlike some other UX Ori stars, the SAGE050756--703453 variations do not seem to be periodic at least over the epochs covered by EROS-2 and MACHO \citep{2005A&A...432..619D}. Figure 3 shows that the star is at maximum light with only some small variations for about 1800 d, followed by two significant declines in brightness over 1300 d and then another 800 d almost constant at maximum. In both the MACHO V-R and EROS-2 V-I colors, the star becomes redder as it fades. The V-R color increases from 0.4 to 0.5-0.6, and V-I color increases from 0.5 to 0.8-0.9 during brightness declines. During the largest decline centered at $\sim$JD 2451700, where SAGE050756--703453 dropped $\Delta$V$\sim$2 mag, the V-I color becomes bluer when the star is deep in the decline.  At first, the color reddens from 0.5 to 0.8 during the initial fading but then becomes bluer again to 0.4 during the last drop to minimum brightness. This behavior is often seen in the declines of UX Ori stars \citep{1998ARA&A..36..233W}. It is caused by preferential scattering of blue light by dust grains around the star back into the beam. This effect becomes important when the stellar photosphere is obscured during the declines.

At maximum light, the colors of SAGE050756--703453 from the MACHO and EROS-2 photometry are B-V$\sim$0.55, V-R$\sim$0.4, and V-I$\sim$0.5. If it is an A2 star, then its intrinsic colors are B-V$\sim$0.05, V-R$\sim$0.03, and V-I$\sim$0.07. The observed B-V and V-R colors are then consistent with an extinction of A$_V\sim$1.5 mag assuming R$_V$=3.1 and a CCM wavelength dependence \citep{1989ApJ...345..245C}. The measured V-I color is too blue but this may be due to the very wide non-standard filters used by EROS-2. In a deep decline two magnitudes below maximum, the measured colors are B-V$\sim$0.7, V-R$\sim$0.5, and V-I$\sim$0.8. The B-V and V-R colors are consistent with an added extinction of only $\sim$0.5 mag for a total of 2 mag, so they are much bluer than expected for a total extinction of 1.5 + 2 = 3.5 mag. Again the EROS-2 V-I color is even bluer. Some of this may be due to the ``blueing" effects of dust scattering or of emission lines. The one set of standard UBVI photometry taken during a 2 mag decline gives the colors, B-V=0.72$\pm$0.05, U-B=$-$0.27$\pm$0.07, V-I= 2.25$\pm$0.09 \citep{2004AJ....128.1606Z}. This V-I color is very red, what would be expected from an extinction of A$_V$$\sim$4 mag. Since this is only one photometric point, it is hard to put it in the context of the lightcurve. It does seem to imply that the blueing effect is not seen in all declines of SAGE050756--703453. 

\begin{figure*}
\figurenum{3a} 
\begin{center}
\includegraphics[width=5in,angle=0]{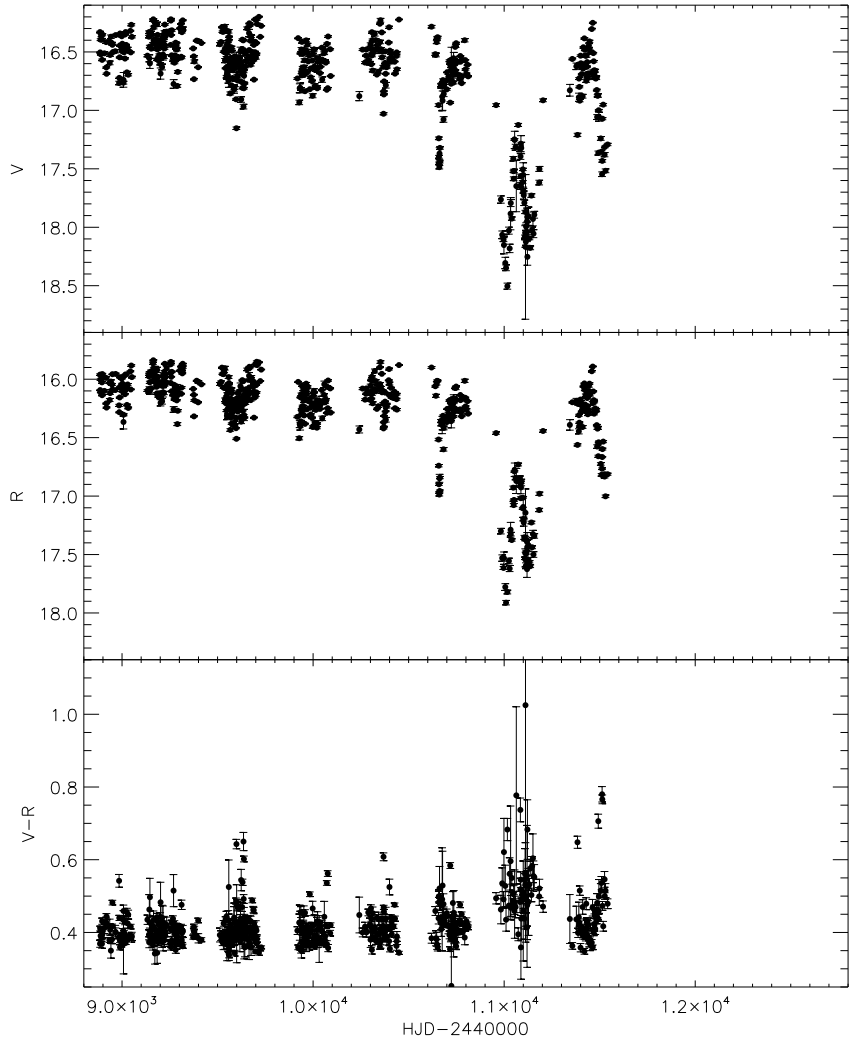}
\end{center}
\caption{MACHO Lightcurve for SAGE050756--703453. The three panels show V, R, and V-R. The MACHO and EROS-2 lightcurves are directly comparable since the range of JD is the same for both plots.}
\end{figure*}

\begin{figure*}
\figurenum{3b} 
\begin{center}
\includegraphics[width=5in,angle=0]{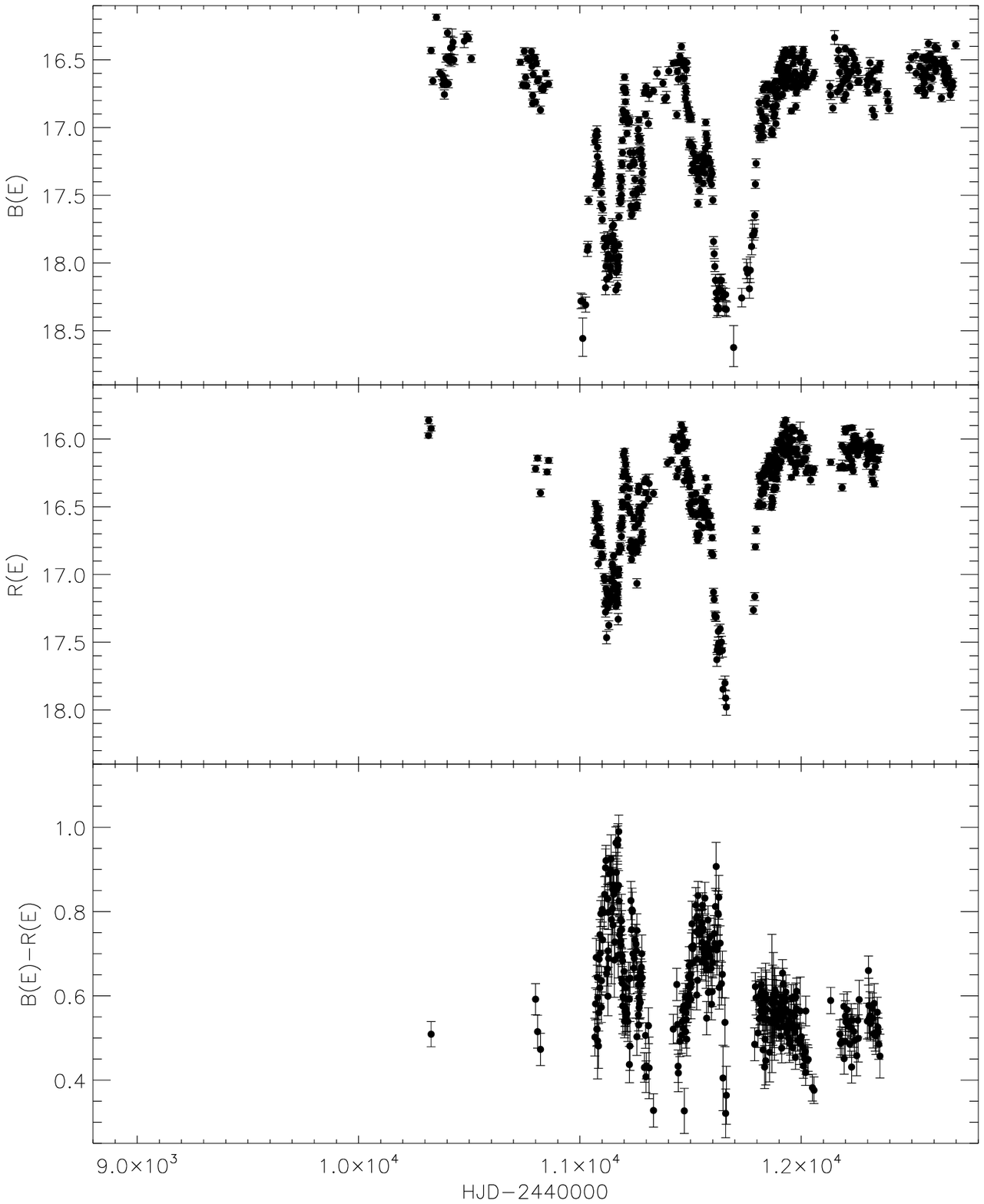}
\end{center}
\caption{EROS-2 lightcurve. The three panels show  B$_E$, R$_E$, and B$_E$ - R$_E$. The MACHO and EROS-2 lightcurves are directly comparable since the range of JD is the same for both plots.}
\end{figure*}

SAGE050756--703453 is V$\sim$16.5 at maximum. If we assume it is a member of the LMC (m-M=18.5) and the extinction is A$_V$=1.5 then M$_V$$\sim$--3.5. Assuming that it is an A2 star (T$_{eff}$=9000 K) implies that R$_{\star}$=21 R$_{\sun}$ and M$_{Bol}$$\sim$-3.8 (L/L$_{\sun}$$\sim$2600). This would make SAGE050756--703453 significantly brighter than Luminosity Class III and brighter than typical Galactic Herbig Ae/Be stars of the same effective temperature \citep{1998A&AS..133...81T}. However, there is an indication from some Herbig Ae/Be star candidates discovered so far in the LMC, that they are brighter than their Galactic counterparts \citep{2005A&A...432..619D}. If the estimated effective temperature and bolometric luminosity of SAGE050756--703453 are correct then it may require a high protostar accretion rate ($\sim10^{-4}$M$_{\sun}$ yr$^{-1}$ \citep{1993ApJ...418..414P}. Obviously, if the star were a foreground object, then it would be much fainter intrinsically but there are no sites of star formation in the Galaxy along the line of sight toward SAGE050756--703453 where a Herbig Ae/Be star would be likely to be found. 
The measured radial velocities of the absorption lines in the SAGE050756--703453 spectrum do not show a redshift of $\sim$200 km s$^{-1}$ typical of an LMC member. But these lines are partially filled in an unknown amount of P-Cygni emission which may account for the lack of redshift. 
It still seems more likely that
SAGE050756--703453 is a member of the LMC.

\begin{figure*}
\figurenum{4} 
\begin{center}

\includegraphics[width=7in,angle=0]{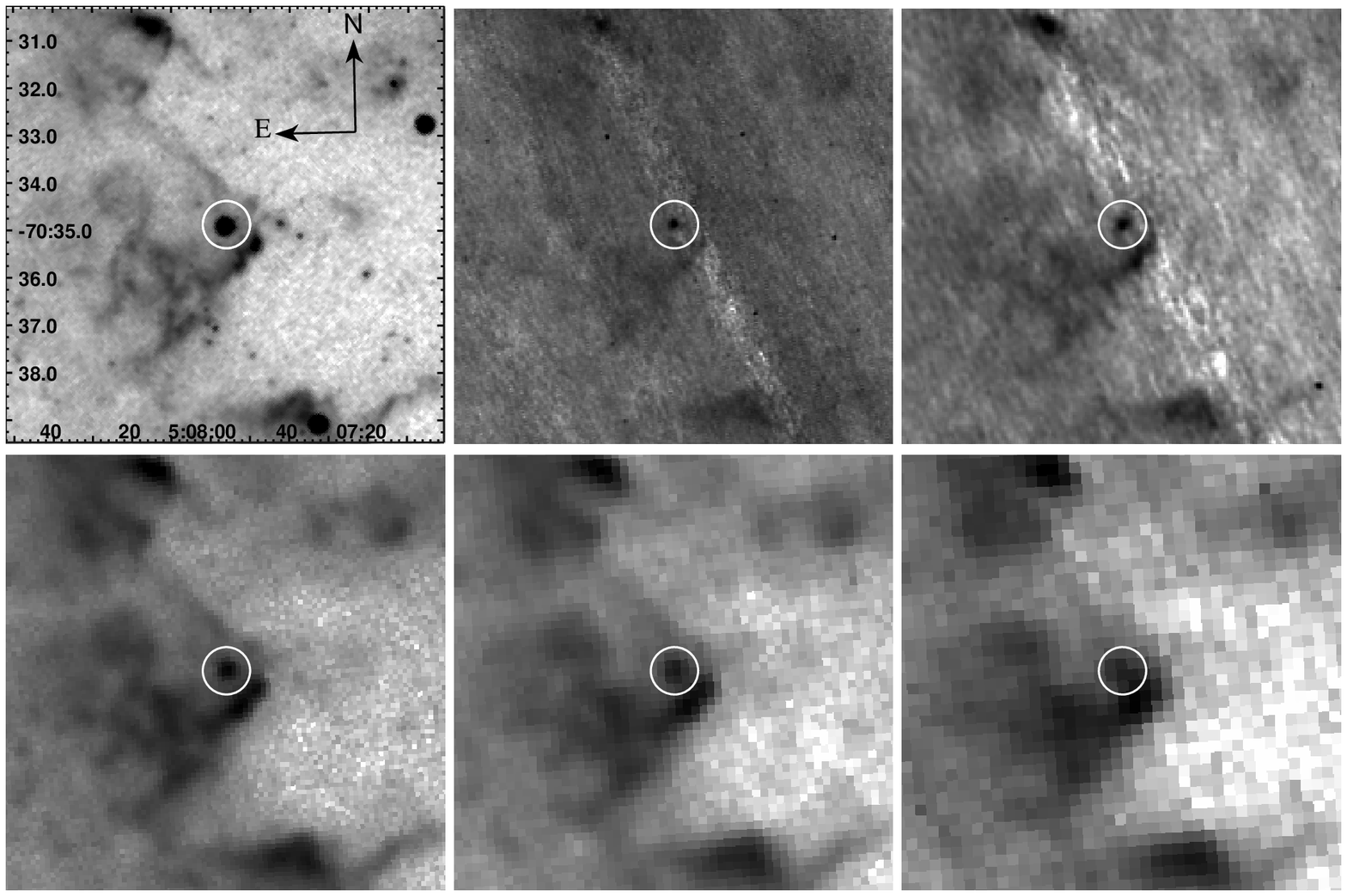}
\end{center}
\caption{Images of the SAGE050756--703453 field. Top row (L to R): MIPS 24 \micron, PACS 100 and 160 \micron. Bottom row (L to R): SPIRE, 250, 350 and 500 \micron. The star is clearly visible as a point source in the center of each field.}
\end{figure*}

SAGE050756--703453 shows extended H$\alpha$ emission on images obtained for the Reid-Parker LMC PN survey \citep{2006MNRAS.365..401R,2006MNRAS.373..521R}. This is a typical characteristic of Herbig Ae/Be stars which excite nearby ISM creating an H II region. Even at the distance of the LMC, an A star can excite an H II region of arcminute size if the gas densities are very low. The Spitzer and Herschel images shown in Figure 4 indicate that SAGE050756--703453 lies in a complex region of dust emission. However, it is clearly a point source at all wavelengths out to 500 \micron~which is consistent with this dust being in a circumstellar disk or envelope.

\citet{1997ApJ...491..885N} modeled the SEDs of 30 Herbig Ae/Be stars, about half of which display the large brightness variations typical of the UX Ori stars. Eight of the stars had measurements of dust continuum emission at 1.3 mm implying circumstellar dust masses of 10$^{-3}$ - 10$^{-5}$ M$_{\sun}$. 
Radiative transfer modeling of IR and millimeter observations of UX Ori, itself, assumed the circumstellar dust was in a flat or flared disk \citep{1999A&A...350..541N}. A reasonable fit was achieved only by including two populations of silicate grains \citep{1984ApJ...285...89D}, one with a size of 1 \micron~to fit the mid-IR, and one with very large ``pebble" sized (10 cm) grains to fit the near-IR and millimeter emission. 

The SED of SAGE050756--703453 is plotted in Figure 5 using the data from Table 1. We have modeled the SED using a Monte Carlo radiation transfer code including nonisotropic scattering, polarization, and thermal emission from dust in a spherical-polar grid \citep{2003ApJ...598.1079W,2003ApJ...591.1049W,2006ApJS..167..256R}. The circumstellar geometry consists of a rotationally flattened infalling envelope, bipolar cavities, and a flared accretion disk. The luminosity sources include the central star and  disk accretion. 
The best fit model includes a disk surrounded by a large diffuse envelope similar to the models used to fit other LMC YSOs with Herschel photometry
\citep{2010A&A...518L..73S}. The SAGE050756--703453 model assumes T$_{eff}$=10000 K,
R$_{\star}$ = 20 R$_{\sun}$, and L$_{\star}$ = 3585 L$_{\sun}$. The disk parameters (mass = 0.02 M$_{\sun}$, r$_{inner}$ = 4.36 AU, r$_{outer}$ = 100 AU) are not well constrained because the SED is dominated by far-IR emission from the envelope. 
The value of L$_{\star}$ = 3585 L$_{\sun}$ is the best model fit to the SED, but this value is not significantly different from the value of 2600 L$_{\sun}$ quoted above within the inherent uncertainties particularly the amount of extinction present in front of the star.
To get a good fit to the far-IR Herschel bands, the envelope must be very large (r$_{inner}$ = 4.36 AU, r$_{outer}$ = 3 pc) but very low density. 
If there is even a small amount of dust in the vicinity of this luminous star, it will be heated even at a distance of a few pc.  Ê
The envelope need not be circumstellar, gravitationally-bound material. It could just be low-density ambient material from the surrounding molecular cloud that is heated both by the internal star and the external interstellar radiation field. The density of the envelope is so low that SAGE050756-70345 would still appear as a point source at LMC distances. 
The best model fit implies a bipolar cavity with an opening angle of 20-40\degree (half-opening angle from pole) and an envelope infall rate of $\sim$2 $\times$ 10$^{-4}$ M$_{\sun}$ yr$^{-1}$.
 In Figure 5, the green dotted SED fit assumes that the cavity is filled with a small amount of dust with a constant density profile. But a much better fit (solid green line) is provided in the IRAC bands by assuming a r$^{-2}$ density profile. ÊThat is consistent with a spherically symmetric outflow, i.e.,  a stellar wind as opposed to a jet which is cylindrical and would be closer to a constant density profile. 
If this is the case, then the light variations seen in Figure 4 might be caused by new dust condensing in the flow \citep{1991PASP..103.1069K}. 
Evidence for outflowing gas in SAGE050756--703453 can be seen in its P-Cygni emission line profiles (Figure 2) but no indication of infall is seen. However, inverse P-Cygni profiles are generally seen only near maximum light \citep{2002ApJ...564..405R}. But the infall rate inferred by this model fits well with the estimated absolute bolometric luminosity of SAGE050756--703453.


Neither of the LMC UX Ori candidates, SAGE050756--703453 nor ELHC7, appear in the \citet{2009ApJS..184..172G} list of YSO candidates because their IRAC [4.5]-[8.0] colors are less than 2.0. However, they appear in the \citet{2008AJ....136...18W} list because different selection criteria were used. 
SAGE050756--703453 does have a larger near-IR excess (H-K=1.42, J-H=1.60) than ELHC7 (H-K=1.03, J-H= 0.61) indicating significant warm dust is present \citep{2005A&A...432..619D}. 
ELHC7 is detected in the IRAC bands, but cannot be seen at longer wavelengths in the MIPS, PACS or SPIRE images. Also, from the MACHO photometry, ELHC7 is fainter and bluer, V$\sim$17.3, V-I$\sim$0.2 so it is likely that it is intrinsically fainter than SAGE050756--703453 \citep{2005A&A...432..619D}. Its brightness variations have a higher frequency and a lower amplitude, and also show evidence for a quasi-period of $\sim$190 d. SAGE050756--703453 shows no sign of periodic brightness variations. 
SAGE050756--703453 has shown two significant declines over $\sim$4000 d
while ELHC7 showed variations of up to $\sim$0.9 mag over a similar timescale.

\begin{figure*}
\figurenum{5} 
\begin{center}
\includegraphics[width=6in,angle=0]{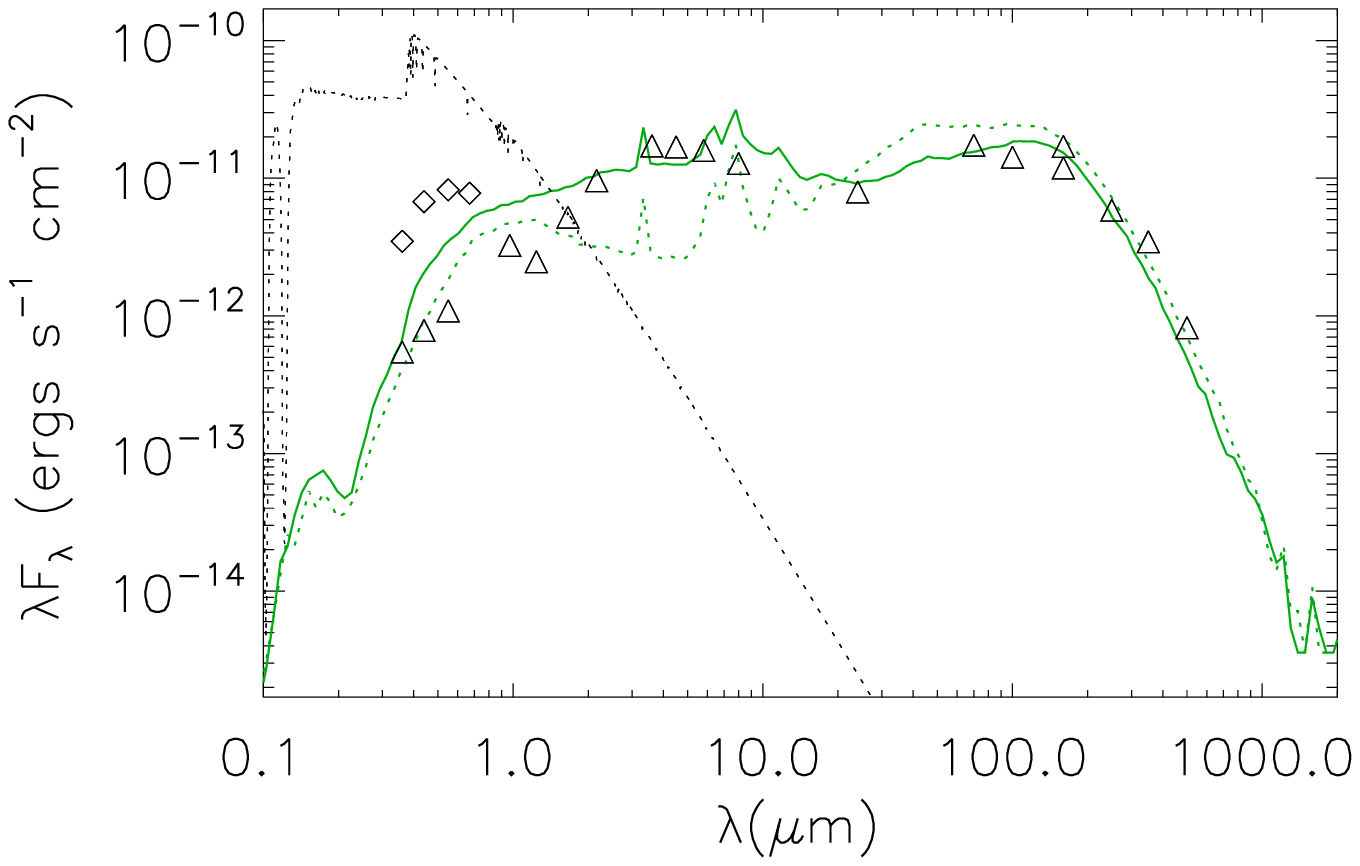}
\end{center}
\caption{SAGE050756--703453 SED including UBVIJHK, IRAC, MIPS, PACS and SPIRE photometry plotted as triangles. Another epoch of UBVR photometry is plotted as diamonds to emphasize that this is a variable object. (See text).
The dotted green line is a model with constant density in an outflow cavity while the solid green line is a better fitting model with the density proportional to r$^{-2}$.  The dotted black line is the input A2-type stellar atmosphere spectrum.}
\end{figure*}

\section{Summary}

SAGE050756--703453 has all of the characteristics of a classic UX Ori star. It has a spectral type of $\sim$A2 and has excited an H II region in its vicinity.  It shows irregular drops in brightness of up to 2 mag and displays the reddening and ``blueing" typical of this class of stars. Its spectrum, showing a combination of emission and absorption lines, is typical of a UX Ori star that is in a decline caused by obscuration from the circumstellar disk. The star has a strong IR excess and significant emission is present out to 500 \micron~similar to that seen in UX Ori, itself 
\citep{1999A&A...350..541N}. However, fits to the SED indicate that in addition to the dusty disk which is needed to explain the visible light variations, dust in bipolar cavities and a very large diffuse envelope are required to fit the SED at all wavelengths. Like other LMC Herbig Ae/Be stars, SAGE050756--703453 is overluminous compared to their Galactic counterparts. Multi-wavelength studies of other LMC PMS stars are needed to determine the reasons for the luminosity differences.

\acknowledgments
We thank the anonymous referee for several useful suggestions that improved the paper. This publication makes use of data products from the Two Micron All Sky Survey, which is a joint project of the University of Massachusetts and the Infrared Processing and Analysis Center/California Institute of Technology, funded by the National Aeronautics and Space Administration and the National Science Foundation. We thank Warren Reid and Quentin Parker for providing access to their LMC H$\alpha$ survey data. 
We acknowledge financial support from the
NASA Herschel Science Center, JPL contract Nos. 1381522 and 1381650. We appreciate the contributions
and support from the European Space Agency (ESA), the PACS and SPIRE
teams, the Herschel Science Center and the NASA Herschel Science Center (esp.\
A. Barbar and K. Xu) and the PACS/SPIRE instrument control center at CEA-Saclay,
which made this work possible.

\bibliography{/Users/gclayton/projects/latexstuff/everything2}

\begin{thebibliography}{45}
\expandafter\ifx\csname natexlab\endcsname\relax\def\natexlab#1{#1}\fi

\bibitem[{{Alcock} {et~al.}(1999){Alcock}, {Allsman}, {Alves}, {Axelrod},
  {Becker}, {Bennett}, {Cook}, {Drake}, {Freeman}, {Geha}, {Griest}, {Lehner},
  {Marshall}, {Minniti}, {Peterson}, {Popowski}, {Pratt}, {Nelson}, {Quinn},
  {Stubbs}, {Sutherland}, {Tomaney}, {Vandehei}, {Welch}, \& {The MACHO
  Collaboration}}]{1999PASP..111.1539A}
{Alcock}, C., {et~al.} 1999, \pasp, 111, 1539

\bibitem[{{Beaulieu} {et~al.}(1996){Beaulieu}, {Lamers}, {Grison}, {Julien},
  {Lanciaux}, {Ferlet}, {Vidal-Madjar}, {Bertin}, {Maurice}, {Prevot}, {Gry},
  {Guibert}, {Moreau}, {Tajhmady}, {Aubourg}, {Bareyre}, {de Kat}, {Gros},
  {Laurent}, {Lachieze-Rey}, {Lesquoy}, {Magneville}, {Milsztajn}, {Moscoso},
  {Queinnec}, {Renault}, {Rich}, {Spiro}, {Vigroux}, {Zylberajch}, {Ansari},
  {Cavalier}, \& {Moniez}}]{1996Sci...272..995B}
{Beaulieu}, J.~P., {et~al.} 1996, Science, 272, 995

\bibitem[{{Beaulieu} {et~al.}(1997){Beaulieu}, {Sasselov}, {Renault}, {Grison},
  {Ferlet}, {Vidal-Madjar}, {Maurice}, {Prevot}, {Aubourg}, {Bareyre},
  {Brehin}, {Coutures}, {Delabrouille}, {de Kat}, {Gros}, {Laurent},
  {Lachieze-Rey}, {Lesquoy}, {Magneville}, {Milsztajn}, {Moscoso}, {Queinnec},
  {Rich}, {Spiro}, {Vigroux}, {Zylberajch}, {Ansari}, {Cavalier}, {Moniez},
  {Gry}, {Guibert}, {Moreau}, \& {Tajhmady}}]{1997A&A...318L..47B}
---. 1997, \aap, 318, L47

\bibitem[{{Bertout}(2000)}]{2000A&A...363..984B}
{Bertout}, C. 2000, \aap, 363, 984

\bibitem[{{Bowen}(1947)}]{1947PASP...59..196B}
{Bowen}, I.~S. 1947, \pasp, 59, 196

\bibitem[{{Brandner} {et~al.}(2001){Brandner}, {Grebel}, {Barb{\'a}},
  {Walborn}, \& {Moneti}}]{2001AJ....122..858B}
{Brandner}, W., {Grebel}, E.~K., {Barb{\'a}}, R.~H., {Walborn}, N.~R., \&
  {Moneti}, A. 2001, \aj, 122, 858

\bibitem[{{Cardelli} {et~al.}(1989){Cardelli}, {Clayton}, \&
  {Mathis}}]{1989ApJ...345..245C}
{Cardelli}, J.~A., {Clayton}, G.~C., \& {Mathis}, J.~S. 1989, \apj, 345, 245

\bibitem[{{Cutri} {et~al.}(2003){Cutri}, {Skrutskie}, {van Dyk}, {Beichman},
  {Carpenter}, {Chester}, {Cambresy}, {Evans}, {Fowler}, {Gizis}, {Howard},
  {Huchra}, {Jarrett}, {Kopan}, {Kirkpatrick}, {Light}, {Marsh}, {McCallon},
  {Schneider}, {Stiening}, {Sykes}, {Weinberg}, {Wheaton}, {Wheelock}, \&
  {Zacarias}}]{2003tmc..book.....C}
{Cutri}, R.~M., {et~al.} 2003, {2MASS All Sky Catalog of point sources.}

\bibitem[{{de Wit} {et~al.}(2003){de Wit}, {Beaulieu}, {Lamers}, {Lesquoy}, \&
  {Marquette}}]{2003A&A...410..199D}
{de Wit}, W.~J., {Beaulieu}, J., {Lamers}, H.~J.~G.~L.~M., {Lesquoy}, E., \&
  {Marquette}, J. 2003, \aap, 410, 199

\bibitem[{{de Wit} {et~al.}(2002){de Wit}, {Beaulieu}, \&
  {Lamers}}]{2002A&A...395..829D}
{de Wit}, W.~J., {Beaulieu}, J.~P., \& {Lamers}, H.~J.~G.~L.~M. 2002, \aap,
  395, 829

\bibitem[{{de Wit} {et~al.}(2005){de Wit}, {Beaulieu}, {Lamers}, {Coutures}, \&
  {Meeus}}]{2005A&A...432..619D}
{de Wit}, W.~J., {Beaulieu}, J.~P., {Lamers}, H.~J.~G.~L.~M., {Coutures}, C.,
  \& {Meeus}, G. 2005, \aap, 432, 619

\bibitem[{{Draine} \& {Lee}(1984)}]{1984ApJ...285...89D}
{Draine}, B.~T., \& {Lee}, H.~M. 1984, \apj, 285, 89

\bibitem[{{Dullemond} {et~al.}(2003){Dullemond}, {van den Ancker}, {Acke}, \&
  {van Boekel}}]{2003ApJ...594L..47D}
{Dullemond}, C.~P., {van den Ancker}, M.~E., {Acke}, B., \& {van Boekel}, R.
  2003, \apjl, 594, L47

\bibitem[{{Finkenzeller}(1985)}]{1985A&A...151..340F}
{Finkenzeller}, U. 1985, \aap, 151, 340

\bibitem[{{Grady} {et~al.}(2000){Grady}, {Sitko}, {Russell}, {Lynch}, {Hanner},
  {Perez}, {Bjorkman}, \& {de Winter}}]{2000prpl.conf..613G}
{Grady}, C.~A., {Sitko}, M.~L., {Russell}, R.~W., {Lynch}, D.~K., {Hanner},
  M.~S., {Perez}, M.~R., {Bjorkman}, K.~S., \& {de Winter}, D. 2000, Protostars
  and Planets IV, 613

\bibitem[{{Gruendl} \& {Chu}(2009)}]{2009ApJS..184..172G}
{Gruendl}, R.~A., \& {Chu}, Y. 2009, \apjs, 184, 172

\bibitem[{{Herbig}(1960)}]{1960ApJS....4..337H}
{Herbig}, G.~H. 1960, \apjs, 4, 337

\bibitem[{{Herbst} {et~al.}(1994){Herbst}, {Herbst}, {Grossman}, \&
  {Weinstein}}]{1994AJ....108.1906H}
{Herbst}, W., {Herbst}, D.~K., {Grossman}, E.~J., \& {Weinstein}, D. 1994, \aj,
  108, 1906

\bibitem[{{Hern{\'a}ndez} {et~al.}(2004){Hern{\'a}ndez}, {Calvet},
  {Brice{\~n}o}, {Hartmann}, \& {Berlind}}]{2004AJ....127.1682H}
{Hern{\'a}ndez}, J., {Calvet}, N., {Brice{\~n}o}, C., {Hartmann}, L., \&
  {Berlind}, P. 2004, \aj, 127, 1682

\bibitem[{{Kenyon} {et~al.}(1991){Kenyon}, {Hartmann}, \&
  {Kolotilov}}]{1991PASP..103.1069K}
{Kenyon}, S.~J., {Hartmann}, L.~W., \& {Kolotilov}, E.~A. 1991, \pasp, 103,
  1069

\bibitem[{{Lamers} {et~al.}(1999){Lamers}, {Beaulieu}, \& {de
  Wit}}]{1999A&A...341..827L}
{Lamers}, H.~J.~G.~L.~M., {Beaulieu}, J.~P., \& {de Wit}, W.~J. 1999, \aap,
  341, 827

\bibitem[{{Lindsay}(1963)}]{1963IrAJ....6..127L}
{Lindsay}, E.~M. 1963, Irish Astronomical Journal, 6, 127

\bibitem[{{Massey}(2002)}]{2002ApJS..141...81M}
{Massey}, P. 2002, \apjs, 141, 81

\bibitem[{{Meixner} {et~al.}(2006){Meixner}, {Gordon}, {Indebetouw}, {Hora},
  {Whitney}, {Blum}, {Reach}, {Bernard}, {Meade}, {Babler}, {Engelbracht},
  {For}, {Misselt}, {Vijh}, {Leitherer}, {Cohen}, {Churchwell}, {Boulanger},
  {Frogel}, {Fukui}, {Gallagher}, {Gorjian}, {Harris}, {Kelly}, {Kawamura},
  {Kim}, {Latter}, {Madden}, {Markwick-Kemper}, {Mizuno}, {Mizuno}, {Mould},
  {Nota}, {Oey}, {Olsen}, {Onishi}, {Paladini}, {Panagia}, {Perez-Gonzalez},
  {Shibai}, {Sato}, {Smith}, {Staveley-Smith}, {Tielens}, {Ueta}, {van Dyk},
  {Volk}, {Werner}, \& {Zaritsky}}]{2006AJ....132.2268M}
{Meixner}, M., {et~al.} 2006, \aj, 132, 2268

\bibitem[{{Meixner} {et~al.}(2010){Meixner}, {Galliano}, {Hony}, {Roman-Duval},
  {Robitaille}, {Panuzzo}, {Sauvage}, {Gordon}, {Engelbracht}, {Misselt},
  {Okumura}, {Beck}, {Bernard}, {Bolatto}, {Bot}, {Boyer}, {Bracker},
  {Carlson}, {Clayton}, {Chen}, {Churchwell}, {Fukui}, {Galametz}, {Hora},
  {Hughes}, {Indebetouw}, {Israel}, {Kawamura}, {Kemper}, {Kim}, {Kwon},
  {Lawton}, {Li}, {Long}, {Marengo}, {Madden}, {Matsuura}, {Oliveira},
  {Onishi}, {Otsuka}, {Paradis}, {Poglitsch}, {Riebel}, {Reach}, {Rubio},
  {Sargent}, {Sewi{\l}o}, {Simon}, {Skibba}, {Smith}, {Srinivasan}, {Tielens},
  {van Loon}, {Whitney}, \& {Woods}}]{2010A&A...518L..71M}
---. 2010, \aap, 518, L71

\bibitem[{{Mendoza}(1971)}]{1971BOTT....6..137M}
{Mendoza}, E.~E. 1971, Boletin de los Observatorios Tonantzintla y Tacubaya, 6,
  137

\bibitem[{{Natta} {et~al.}(1997){Natta}, {Grinin}, {Mannings}, \&
  {Ungerechts}}]{1997ApJ...491..885N}
{Natta}, A., {Grinin}, V.~P., {Mannings}, V., \& {Ungerechts}, H. 1997, \apj,
  491, 885

\bibitem[{{Natta} {et~al.}(1999){Natta}, {Prusti}, {Neri}, {Thi}, {Grinin}, \&
  {Mannings}}]{1999A&A...350..541N}
{Natta}, A., {Prusti}, T., {Neri}, R., {Thi}, W.~F., {Grinin}, V.~P., \&
  {Mannings}, V. 1999, \aap, 350, 541

\bibitem[{{Natta} \& {Whitney}(2000)}]{2000A&A...364..633N}
{Natta}, A., \& {Whitney}, B.~A. 2000, \aap, 364, 633

\bibitem[{{Palla} \& {Stahler}(1993)}]{1993ApJ...418..414P}
{Palla}, F., \& {Stahler}, S.~W. 1993, \apj, 418, 414

\bibitem[{{Reid} \& {Parker}(2006{\natexlab{a}})}]{2006MNRAS.365..401R}
{Reid}, W.~A., \& {Parker}, Q.~A. 2006{\natexlab{a}}, \mnras, 365, 401

\bibitem[{{Reid} \& {Parker}(2006{\natexlab{b}})}]{2006MNRAS.373..521R}
---. 2006{\natexlab{b}}, \mnras, 373, 521

\bibitem[{{Robitaille} {et~al.}(2006){Robitaille}, {Whitney}, {Indebetouw},
  {Wood}, \& {Denzmore}}]{2006ApJS..167..256R}
{Robitaille}, T.~P., {Whitney}, B.~A., {Indebetouw}, R., {Wood}, K., \&
  {Denzmore}, P. 2006, \apjs, 167, 256

\bibitem[{{Rodgers} {et~al.}(1988){Rodgers}, {Conroy}, \&
  {Bloxham}}]{1988PASP..100..626R}
{Rodgers}, A.~W., {Conroy}, P., \& {Bloxham}, G. 1988, \pasp, 100, 626

\bibitem[{{Rodgers} {et~al.}(2002){Rodgers}, {Wooden}, {Grinin}, {Shakhovsky},
  \& {Natta}}]{2002ApJ...564..405R}
{Rodgers}, B., {Wooden}, D.~H., {Grinin}, V., {Shakhovsky}, D., \& {Natta}, A.
  2002, \apj, 564, 405

\bibitem[{{Seale} {et~al.}(2009){Seale}, {Looney}, {Chu}, {Gruendl}, {Brandl},
  {Chen}, {Brandner}, \& {Blake}}]{2009ApJ...699..150S}
{Seale}, J.~P., {Looney}, L.~W., {Chu}, Y., {Gruendl}, R.~A., {Brandl}, B.,
  {Chen}, C., {Brandner}, W., \& {Blake}, G.~A. 2009, \apj, 699, 150

\bibitem[{{Sewi{\l}o} {et~al.}(2010){Sewi{\l}o}, {Indebetouw}, {Carlson},
  {Whitney}, {Chen}, {Meixner}, {Robitaille}, {van Loon}, {Oliveira},
  {Churchwell}, {Simon}, {Hony}, {Panuzzo}, {Sauvage}, {Roman-Duval}, {Gordon},
  {Engelbracht}, {Misselt}, {Okumura}, {Beck}, {Hora}, \&
  {Woods}}]{2010A&A...518L..73S}
{Sewi{\l}o}, M., {et~al.} 2010, \aap, 518, L73

\bibitem[{{Testi} {et~al.}(1998){Testi}, {Palla}, \&
  {Natta}}]{1998A&AS..133...81T}
{Testi}, L., {Palla}, F., \& {Natta}, A. 1998, \aaps, 133, 81

\bibitem[{{Tisserand} {et~al.}(2009){Tisserand}, {Wood}, {Marquette}, {Afonso},
  {Albert}, {Andersen}, {Ansari}, {Aubourg}, {Bareyre}, {Beaulieu}, {Charlot},
  {Coutures}, {Ferlet}, {Fouqu{\'e}}, {Glicenstein}, {Goldman}, {Gould},
  {Gros}, {de Kat}, {Lesquoy}, {Loup}, {Magneville}, {Maurice}, {Maury},
  {Milsztajn}, {Moniez}, {Palanque-Delabrouille}, {Perdereau}, {Rich},
  {Schwemling}, {Spiro}, \& {Vidal-Madjar}}]{2009A&A...501..985T}
{Tisserand}, P., {et~al.} 2009, \aap, 501, 985

\bibitem[{{Vijh} {et~al.}(2009){Vijh}, {Meixner}, {Babler}, {Block}, {Bracker},
  {Engelbracht}, {For}, {Gordon}, {Hora}, {Indebetouw}, {Leitherer}, {Meade},
  {Misselt}, {Sewilo}, {Srinivasan}, \& {Whitney}}]{2009AJ....137.3139V}
{Vijh}, U.~P., {et~al.} 2009, \aj, 137, 3139

\bibitem[{{Waters} \& {Waelkens}(1998)}]{1998ARA&A..36..233W}
{Waters}, L.~B.~F.~M., \& {Waelkens}, C. 1998, \araa, 36, 233

\bibitem[{{Whitney} {et~al.}(2003{\natexlab{a}}){Whitney}, {Wood}, {Bjorkman},
  \& {Cohen}}]{2003ApJ...598.1079W}
{Whitney}, B.~A., {Wood}, K., {Bjorkman}, J.~E., \& {Cohen}, M.
  2003{\natexlab{a}}, \apj, 598, 1079

\bibitem[{{Whitney} {et~al.}(2003{\natexlab{b}}){Whitney}, {Wood}, {Bjorkman},
  \& {Wolff}}]{2003ApJ...591.1049W}
{Whitney}, B.~A., {Wood}, K., {Bjorkman}, J.~E., \& {Wolff}, M.~J.
  2003{\natexlab{b}}, \apj, 591, 1049

\bibitem[{{Whitney} {et~al.}(2008){Whitney}, {Sewilo}, {Indebetouw},
  {Robitaille}, {Meixner}, {Gordon}, {Meade}, {Babler}, {Harris}, {Hora},
  {Bracker}, {Povich}, {Churchwell}, {Engelbracht}, {For}, {Block}, {Misselt},
  {Vijh}, {Leitherer}, {Kawamura}, {Blum}, {Cohen}, {Fukui}, {Mizuno},
  {Mizuno}, {Srinivasan}, {Tielens}, {Volk}, {Bernard}, {Boulanger}, {Frogel},
  {Gallagher}, {Gorjian}, {Kelly}, {Latter}, {Madden}, {Kemper}, {Mould},
  {Nota}, {Oey}, {Olsen}, {Onishi}, {Paladini}, {Panagia}, {Perez-Gonzalez},
  {Reach}, {Shibai}, {Sato}, {Smith}, {Staveley-Smith}, {Ueta}, {Van Dyk},
  {Werner}, {Wolff}, \& {Zaritsky}}]{2008AJ....136...18W}
{Whitney}, B.~A., {et~al.} 2008, \aj, 136, 18

\bibitem[{{Zaritsky} {et~al.}(2004){Zaritsky}, {Harris}, {Thompson}, \&
  {Grebel}}]{2004AJ....128.1606Z}
{Zaritsky}, D., {Harris}, J., {Thompson}, I.~B., \& {Grebel}, E.~K. 2004, \aj,
  128, 1606

\end{thebibliography}


\end{document}